\documentclass[10pt,preprint2]{aastex}

\begin{document}

\title{Multiwavelength Observations of Radio Galaxy 3C 120 with XMM-Newton}

\author{P. M. Ogle$^1$, S. W. Davis$^2$, R. R. J. Antonucci$^2$, J. W. Colbert$^3$, M. A. Malkan$^4$, M. J. Page$^5$, T. P. Sasseen$^2$, M. Tornikoski$^6$}

\affil{$^1$ Jet Propulsion Lab, Mail Code 238-332, 4800 Oak Grove Dr., Pasadena, CA 91109}
\affil{$^2$ Physics Dept., University of California, Santa Barbara, CA 93106}
\affil{$^3$ Spitzer Science Center, Mail Code 220-6, Pasadena, CA 91125}
\affil{$^4$ Astronomy Division, University of California, Los Angeles, CA 90095}
\affil{$^5$ MSSL, University College London, Holmbury St. Mary, Dorking, 
            Surrey, RH5 6NT, UK}
\affil{$^6$ Mets\"{a}hovi Radio Observatory, Helsinki University of Technology, Mets\"{a}hovintie 114, 02540, Kylm\"{a}l\"{a}, 
       Finland}

\email{Patrick.M.Ogle@jpl.nasa.gov}

\shorttitle{Radio Galaxy 3C 120}
\shortauthors{Ogle et al.}

\begin{abstract}

We present {\it XMM-Newton} observations of the radio galaxy 3C 120, which we use to study the nature and
geometry of the X-ray and UV-emitting regions. Contemporaneous radio, mm-wave, and optical data provide
additional constraints on the spectral energy distribution and physical state of the active galactic nucleus. The 
hard X-ray spectrum contains a marginally resolved Fe {\sc i} K$\alpha$ emission line with FWHM$=9,000\pm 3,000$ km s$^{-1}$ 
and an equivalent width of $57\pm 7$ eV. The line arises via fluorescence in a broad-line region with covering
fraction of 0.4. There is no evidence of relativistically broad Fe K$\alpha$, contrary to some 
previous reports. The normal equivalent widths of the X-ray and optical emission lines exclude a strongly beamed 
synchrotron component to the hard X-ray and optical continua.  There is an excess of 0.3-2 keV soft X-ray continuum over 
an extrapolation of the hard X-ray power-law, which may arise in a disk corona. Analysis of an archival {\it Chandra} image
shows that extended emission from the jet and other sources contributes $<3\%$ of the total X-ray flux. A break in the X-ray spectrum 
below 0.6 keV indicates an excess neutral hydrogen column density of $N_\mathrm{H}=1.57 \pm 0.03 \times 10^{21}$ cm$^{-2}$. However, 
the neutral absorber must have an oxygen abundance of $<1/50$ of the solar value to explain the absence of an intrinsic or intervening O {\sc i} 
edge. There is no ionized absorption in the soft X-ray spectrum, but there is a weak, narrow O {\sc viii} Ly$\alpha$ 
emission line. We do not detect previously claimed O {\sc viii} absorption from the intervening intergalactic medium. Radio 
observations at 37 GHz show a fast, high frequency flare starting 7 days after the {\it XMM-Newton} observation. However, this 
has no obvious effect on the X-ray spectrum. The X-ray spectrum, including the soft excess, became harder as the X-ray flux 
decreased, with an estimated pivot energy of $40$ keV. The UV and soft X-ray fluxes are strongly correlated over the 120 ks duration 
of the {\it XMM-Newton} observation. This is qualitatively consistent with Comptonization of UV photons by a hot corona.

\end{abstract}

   
\section{Introduction}

3C 120 (z=0.033) is one of the brightest nearby radio galaxies, classified as a broad-line 
radio galaxy (BLRG). It belongs to the Fanaroff-Riley (FR) I class of radio galaxies, which generally have lower-power 
radio jets than FR II sources. The strong broad emission lines are unusual for an FR I radio galaxy. 
3C 120 has a one-sided superluminal jet \citep{s79} with complex, possibly helical structure \citep{gma00,h01,wbu01}. 
Components are ejected with apparent motions of 4-6c, indicating a Doppler factor of $\sim 3$ and
and a line-of-sight $<20\arcdeg$ from the jet axis. 3C 120 is one of the first sources discovered to have 
X-ray synchrotron emission coincident with the extended radio jet \citep{hhs99}. The host galaxy is a spiral \citep{s67}, 
which is unusual for a BLRG. The reverberation mass of the central black hole is $\sim3\times 10^7 M_\odot$ \citep{wpm99,pwb98}. 

BLRGs appear to have weaker Compton reflection and weaker Fe K$\alpha$ emission in their 
hard X-ray spectra than radio-quiet Seyfert galaxies \citep{wzs98,g99,exm00}. The reflection fraction in 
3C 120 is $\Omega/2\pi\sim 0.5$, measured with {\it Rossi X-ray Timing Explorer (RXTE)} and {\it Beppo-SAX} 
\citep{exm00, zg01}. Possible reasons for a small  reflection fraction include 1) dilution by X-ray emission 
from a beamed jet, 2) intrinsic differences between accretion disk or torus structure in radio-loud and 
radio-quiet active galactic nuclei \citep{exm00}, and 3) ionization of the disk surface \citep{brf02}. 
 
The profile of the Fe K$\alpha$ line can be used to probe the structure of the inner accretion disk
in some active galactic nuclei (AGN). The spectra of the radio-quiet narrow-line Seyfert 1 Galaxies MCG-6-30-15, 
Mrk 766, and NGC 4051 contain relativistically broadened Fe K$\alpha$ \citep{tnf95,wrb01,pmc01,s03}. Similarly 
broadened K-shell lines of C, N, and O have been reported in the soft X-ray spectra of these Seyferts 
\citep{br01,mbo03,omp04}. A number of other broad-line Seyfert 1s observed with ASCA  may also have relativistically 
broadened Fe K$\alpha$ lines in their spectra \citep{n97}. An important question to resolve is whether any radio-loud 
AGN have relativistically broadened lines. If so, this could be used to compare their accretion disk structure to 
radio quiet AGN. 

3C 120 was observed with {\it ASCA} in 1994 \citep{gsm97}. A broad Fe K$\alpha$ line was found and
fit by a Gaussian with $\sigma=0.8$ keV and equivalent width (EW) of $\sim 400$ eV. The same data were
fit by \cite{r97} and \cite{s99}, also yielding a broad line with very large equivalent width.
The large velocity width is suggestive of Doppler broadening in the inner regions of an accretion disk. 
However, the line parameters are sensitive to the underlying continuum shape and amount of Compton reflection. 
The line width and strength decrease dramatically if the model continuum reflection fraction is increased 
\citep{gsm97} or if the continuum is fit with a broken power law \citep{wzs98}.

Observations of the X-ray continuum show that the spectral hardness is anti-correlated with the X-ray flux.
\citep{h85}. Sometimes the X-ray continuum pivots at an energy of $\sim2$ keV, while at other times the pivot 
energy must be at $>6$ keV \citep{mcf91}. Coordinated multi-wavelength observations hint that 
the hard X-ray slope steepens with increased UV continuum flux, consistent with Comptonization of disk emission 
by thermal electrons \citep{mcf91}. {\it Compton Gamma Ray Observatory} OSSE observations indicate a spectral cutoff 
or break at 100-200 keV, consistent with a very hot corona \citep{wzs98}. 

A soft X-ray excess is found in {\it ROSAT} and {\it Beppo-SAX} spectra \citep{gsm97, zg01}, and can be fit by 
either a broken power law or optically thin thermal emission.  Soft excess emission has been observed in other BLRGs 
\citep{wzs98} and Seyfert 1 galaxies, and its nature is currently under debate. In most cases, where not resolved into
emission lines, it is attributed to the Comptonized tail of thermal emission from an accretion disk. For 3C 120,
\cite{zg01} suggest that the soft excess consists of emission lines from a collisionally ionized plasma in an
extended halo surrounding the host galaxy. This might explain a break in the spectrum below 1 keV. Alternatively,
there may be excess neutral absorption over the nominal Galactic amount \citep{gsm97}.

3C 120 was the target of an extended Very Long Baseline Interferometry (VLBI) and {\it RXTE} monitoring campaign 
from 1997-2000 \citep{mjg02}. Ejection of VLBI jet components was preceded by dips in the 2.4-20 keV X-ray flux and 
hardening of the spectral index. The mean separation between each dip and jet ejection event was 0.1 yr, with 
a low probability of a random association. These dips may result from changes in the inner 
accretion flow and corona geometry associated with ejection of radio jet components. A similar 
phenomenon is seen in micro-quasar GRS 1915+105 \citep{mr94}, where IR-radio flares are preceded by
hard X-ray dips \citep{emm98, mr98}. The exact cause of X-ray dips is unknown, but they must be closely connected
to both jet component ejection and the state of the X-ray emitting region.

We use the {\it XMM-Newton} European Photon Imaging Camera (EPIC) and the Reflection Grating Spectrometers (RGS) to measure 
the X-ray continuum and Fe K$\alpha$ emission line properties of 3C 120. These observations are 
supported by simultaneous UV photometry with the Optical Monitor (OM) and contemporaneous observations of radio variability 
and the optical emission line spectrum. Archival {\it Hubble} optical and {\it Chandra} X-ray images are used to
constrain possible extended emission components in the spectrum. After presenting the observations, we discuss 
their implications for the structure of the X-ray and UV emission regions. 

\section{Observations}

We observed 3C 120 for nearly a full orbit (130 ks) with {\it XMM-Newton} on 26-27 August 2003 (2003.65), starting at 
MJD 52877.233333. The X-ray background was low except for a 10 ks background flare at the end of the observation. We exclude 
this period of high background from our analysis. The Medium optical blocking filter was used to avoid optical contamination 
of the EPIC spectra. The EPIC pn and MOS2 (Metal-Oxide-Silicon) detector observations were taken in Small Window mode to 
minimize pileup effects. MOS1 observations were taken in timing mode, and we disregard them because of calibration difficulties.

All {\it XMM-Newton} data were reduced using the Science Analysis System (SAS) 5.4.1 and event files were reprocessed with the 
tools {\it epchain}, {\it emchain}, and {\it rgsproc}. The first- and second-order RGS1 and RGS2 spectra were added together. 
The RGS response matrices were combined and corrected for small errors in the effective area calibration, using Mrk 421 as a 
standard featureless source. 

We include single and double pixel events in our EPIC analysis. Spectra were extracted from a 
$35\arcsec$ circular aperture centered on the X-ray peak in both the pn and MOS2 images. Backgrounds were estimated using 
regions on the detector away from the source. The net count rates were 20.4 and 6.1 ph s$^{-1}$, respectively, indicating 
insignificant pileup in pn, and moderate pileup in MOS2. We used the SAS tool {\it epatplot} to compare the ratios of single, 
double, and multiple pixel events to model predictions. Pileup was manifest in the MOS2 spectrum as a 10\% deficit at 0.2-0.3 
keV and a 5\% excess at 0.5-10 keV. We re-extracted the MOS2 spectrum, excluding the inner $5\arcsec$ of the point-spread 
function. This reduced the net count rate to 3.7 ph s$^{-1}$, and the pileup fraction to $<2\%$. We use this un-piled MOS2 
spectrum in our analysis below. 

The EPIC pn X-ray light curves (Fig. 1a,b) demonstrate significant variability. The 0.3-2 keV (soft) and
2-10 keV (hard) X-ray count rates decrease gradually from beginning to end of the observation. 
There is a slight increase in the hard count rate $\sim 70$ ks after the start of the observation, then 
it resumes its decline. This is accompanied by an increase in the hardness ratio. We concentrate our initial analysis on 
the total flux spectra for the entire observation. Then we split the EPIC pn spectrum into two 60 ks halves to study the form 
of the spectral variability.

\begin{figure}[t]
\plotone{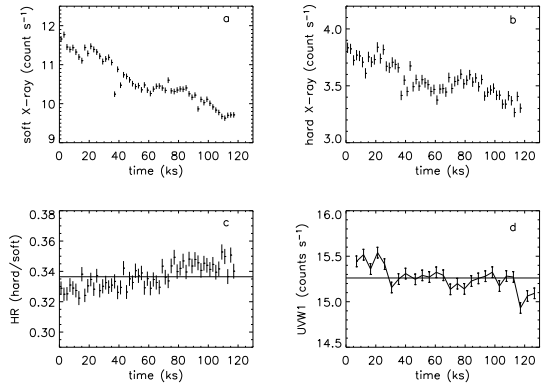}
\figcaption{{\it XMM-Newton} EPIC pn X-ray and OM UV light curves. Background count rates have been subtracted. 
            Time is plotted from the beginning of the observation. 
            The time interval with bright background flare is removed from the end of the observation.
            (a) Soft (0.3-2 keV) count rate. (b) Hard (2-10 keV) count rate. (c) Hardness ratio 
               HR $=$ (2-10 keV count rate)/(0.3-2 keV count rate). (d) OM UVW1 (2900 \AA) light curve.} 
\end{figure}

The {\it XMM-Newton} Optical Monitor \citep{mbm01} was used to measure the UV light curve through the UVW1 
filter (2900 \AA, Fig. 1d), simultaneously with the X-ray observations. OM data were reduced using
the standard {\it omichain} pipeline, with dead time and exposure time corrections applied. Fluxes were measured 
in a square aperture $12\arcsec$ on a side which was centered on the galactic nucleus. The zero-points for the 
UVW1 and V magnitudes are from an XMM calibration document\footnote{
\url{http://xmm.vilspa.esa.es/docs/documents/CAL-TN-0019}}. The flux in the UVW1 band decreased by $2\%$ over 
the course of the observation. The mean UVW1 count rate is 15.26 count s$^{-1}$, yielding a monochromatic 
magnitude of 14.34. There is a small contribution from the Mg II $\lambda 2795$ broad emission line in this band, 
with the EW historically ranging from 37-53 \AA\ \citep{mcf91}. Assuming a similar EW, we estimate that this line
contributes at most 7\% of the observed UVW1 count rate. One V-band exposure was also taken to help pin down the 
SED, giving a count rate of $23.11\pm 0.08$ count s$^{-1}$ and $\mathrm{V}=14.55$ mag. The (uncorrected) 
V-UVW1 spectral index is $\alpha =-0.3$ ($F_\nu\sim \nu^{-\alpha}$).

We made Keck Low-Resolution Imaging Spectrometer (LRIS) observations on 28 August 2003 (MJD 52879.625414), 2 days 
after the {\it XMM-Newton} observation, to measure the optical continuum and broad emission lines. The exposure times 
are 500 s for the blue camera and 400 s for the red camera. Data reductions are made with IRAF. Primary calibrations 
include bias subtraction and flat-fielding. Blue flat-fields are from twilight sky observations, red flat-fields from 
the inside of the illuminated observatory dome. Blue and red spectra are split at 6800 \AA\ by the 680 dichroic, but the 
flux is well-matched across the split. Spectra are extracted using a $1\farcs 3$ aperture. The red wavelength 
scale is calibrated using night sky emission lines, the blue wavelength scale using a combination of sky lines 
and Hg/Cd arc-lamp exposures. Flux calibrations for the red and blue spectra are from the standard star G191B2B, 
including a correction for differential airmass between the source and standard. The monochromatic magnitude at 
5500 \AA\ from the Keck spectrum matches the OM V-band measurement to within 0.1 mag. Similarly, the extrapolation 
of continuum in the Keck spectrum down to  2900 \AA\ is in rough agreement with the UVW1 flux.

\begin{figure}[t]
\plotone{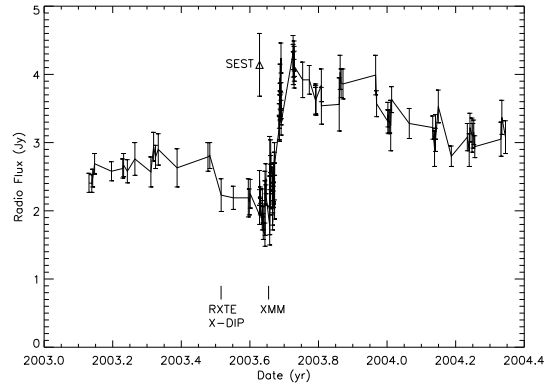}
\figcaption{37 GHz radio flux density. The {\it XMM} observation occurred at 2003.65, just prior 
            to the onset of a large 37 GHz flare. The 250 GHz (1.2 mm) SEST observation is indicated with a 
            triangle.  An X-ray dip observed with RXTE \citep{mja04} was a precursor to the radio flare.}
\end{figure} 

3C 120 was  monitored over the period 17 Feb 2003 - 6 May 2004 (Fig. 2) at Mets\"{a}hovi Radio Observatory 
in Kylm\"{a}l\"{a}, Finland. Observations were made with the 14 m telescope at 37 GHz (8.1 mm). The 37 GHz 
receiver is a dual horn, Dicke-switched receiver with a HEMT preamplifier and is operated at room temperature. 
We use DR 21 as a primary calibrator, against which the flux density is calibrated to the \cite{b77} scale. 
For more details about the Mets\"{a}hovi observing system see \cite{ttm98}. The flux at 37 GHz ranged from 
1.7-4.3 Jy. The estimated flux on the date of the XMM observation was $2.1\pm 0.3$ Jy, interpolated between the 
flux points measured for 25 and 29 August 2003. The source flared from 3-10 September (2003.67-69), doubling its 
radio flux from 2.0-4.2 Jy in 7 days. Fortuitously, this flare occurred only 8 days after the XMM observation. In 
the following months, the flare slowly decayed, with the 37 GHz flux remaining at a relatively high level of 3.1 Jy 
as late as 2004 May.

The source was also observed in the mm band with the 15-meter Swedish-ESO Submillimeter Telescope (SEST) on
Cerro La Silla, Chile. Observations were made on 18 August 2003 by Dr. Y. Chin, 8 days before the 
{\it XMM-Newton} observation. Three sequential observations (spaced 10 min apart) were taken in fast mapping mode
with SIMBA, a 37-channel bolometer array centered at 250 GHz (1.2 mm). For flux calibration Uranus was observed, 
also in fast mapping mode, and opacity corrections were made using frequent sky-dip observations. The maps are 
reduced using the standard reduction package MOPSI\footnote{Created by Robert Zylka at IRAM (Grenoble, France).}. 

The mean 250 GHz flux was $4.13\pm 0.46$ Jy (Table 1), including the uncertainty in the absolute calibration. 
The 37 GHz flux  was $2.2 \pm 0.3$ Jy on the date of the SEST observation, yielding an inverted 37-250 GHz spectral
index of $\alpha=-0.3\pm 0.1$ It is likely that a flare was already underway at 250 GHz at least 17 days before the 
start of the 37 GHz flare, and prior to the {\it XMM-Newton} observation (Fig. 2). The 250 GHz flux was at an unprecedented, 
very high level compared to a historical mean of $\sim 1$ Jy \citep{sps93}. The outburst is confirmed by a December 2003 
observation with the IRAM 30-m radio telescope, when the 150 GHz (2 mm) flux was still high but had dropped to 3.4 
Jy\footnote{Data from the IRAM pointing source catalog at \url{http://www.iram.es/IRAMES/index.htm}}. 

Throughout this paper, we use neutral atomic photoelectric absorption cross sections from \cite{hgd93} and solar abundances from 
\cite{d00}. We use the IMP\footnote{\url{http://xmmom.physics.ucsb.edu/\~\ pmo/imp.html}} and 
XSPEC\footnote{\url{http://heasarc.gsfc.nasa.gov/docs/xanadu/xspec/}.} software packages for our spectral fits, and quote 
90\% confidence intervals for parameter uncertainties. 

\section{ Archival Images}

We analyze an archival Chandra image to evaluate extended emission which might contaminate the XMM spectra of 
the nucleus, and to test the hypothesis that the soft excess is from extended hot gas. A 60 ks HETGS exposure was taken in 
2001 \citep{mym03}. We extracted the 0-order image with CIAO 3.0.2, including events in the full 0.2-12 keV range of the 
detector. The X-ray contours are plotted over an archival HST WF/PC2 image (Fig. 3). In addition to the bright nuclear 
source, two strong X-ray knots are apparent in the image. A detailed analysis of this image and the extended X-ray jet is 
given by \cite{hmw04}. The k25 knot at $25\arcsec$ from the nucleus has a flux density of $1.3\times 10^{-2}$ $\mu$Jy at 1 keV, 
which is only 0.3\% of the 1 keV flux density in our XMM spectra. The k4 knot $4\arcsec$ from the nucleus has a similarly low 
flux of $1.0\times 10^{-2}$ $\mu$Jy at 1 keV.

To search for additional extended emission, we plot the 0.3-2 keV radial point-spread function (PSF) of the Chandra image 
(Fig. 4). For comparison, we plot a model of the Chandra High Resolution Mirror Assembly (HRMA) PSF, from a calibration
file computed using the SAOSAC ray-trace program \footnote{\url{http://cxc.harvard.edu/cal/Hrma/psf/}}. Pileup losses are evident 
at small radii as a deficit in the observed PSF. The peak count rate is 0.28 counts pixel$^{-1}$ frame$^{-1}$, corresponding to 
moderately heavy pileup in the central pixels. The pileup fraction is reduced to $<4\%$ at radii $>0.8\arcsec$. The model PSF 
matches the observation well at $r=6-20\arcsec$. The k4 and k25 knots are evident as excess emission above the model. There also 
appears to be excess emission from $r=1-3\arcsec$, which amounts to 558 counts, or 1.9\% of the total 0.3-2 keV source counts in the
$r=35\arcsec$ aperture, after correcting for pileup. We conclude that extended emission from the jet and other sources 
contributes $<3\%$ of the total flux, and does not contribute significantly to the soft X-ray excess. \cite{hmw04} also find
extended flux at $\sim 0.5-1\arcsec$ in this image, after subtracting a 2D PSF model. Since this emission does not correspond to the 
radio jet, it may be an indicator of starburst activity in the inner disk.

The HST WF/PC2 image was taken 1995 July 25 (Fig. 3). The image is the sum of two $1100$ s exposures in the F675W continuum 
filter (which includes the H$\alpha$ line). We used IRAF/STSDAS to produce a mosaic from the 4 cameras and filter out 
cosmic ray events. The seams between the cameras show up white in the negative image. The host galaxy has a disturbed spiral 
morphology in the outer regions, with many discrete clumps suggestive of star-formation.  There is a faint figure-8 structure 
with lobes stretching ESE-WNW, which is also apparent in the ground-based image taken by \cite{hvs95}, who identify the western 
branch of this feature with optical emission from the jet. However, this connection is questionable, since the other branches 
of the figure-8 do not correspond to emission from the jet. Instead, we suggest that this structure is the signature of a 
late-stage galaxy merger. The bright NW branch of the figure-8 is dominated by continuum emission \citep{hvs95}, consistent 
with starlight. The overexposed inner region of the galaxy has a strong spiral pattern including dust lanes, which winds in 
the opposite sense to the outer spiral arms \citep{hmw04}, another indication of a past merger.

\begin{figure}[t]
\plotone{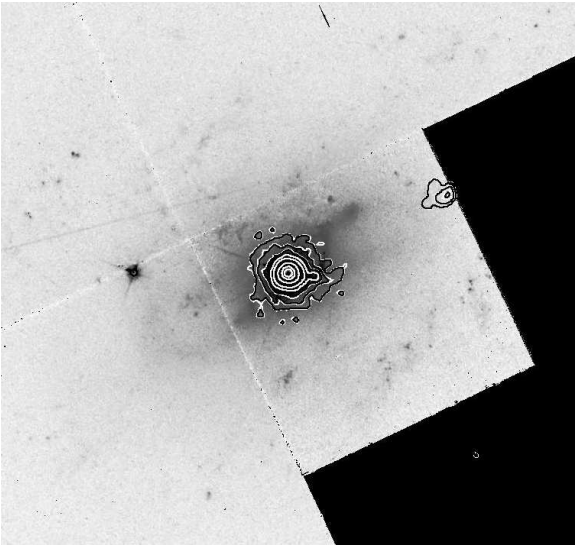}
\figcaption{{\it HST} WF/PC2 F675W image of 3C 120, with Chandra HETGS 0.2-12 keV contours ($69\arcsec \times 69\arcsec$). The 
             outer regions of the galaxy have a disturbed spiral morphology. The figure-8 feature may be tidal debris 
             unrelated to the radio jet. The k4 and k25 X-ray jet components are visible at $4\arcsec$ and 
             $25\arcsec$ from the core. Contours: $1.2\times10^{-3}$, $2.5\times 10^{-3}$, $5\times 10^{-3}$, $1\times 10^{-2}$, 
             $2 \times 10^{-2}$, $8\times 10^{-2}$, 0.32, $0.64 \times$ peak flux. }
\end{figure}

\begin{figure}[t]
\plotone{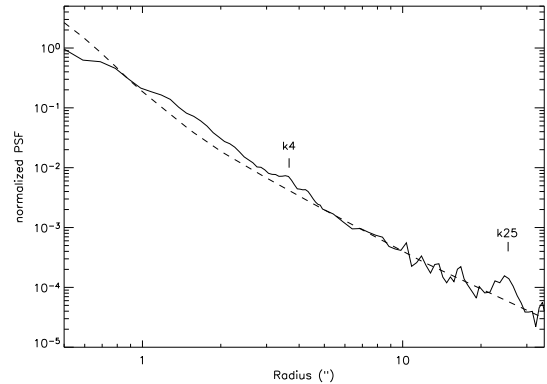}
\figcaption{The radial profile of the {\it Chandra} HETGS image (solid line) is compared to a model of the Chandra point-spread
            function (dashed line). The peak of the observed profile is reduced by pileup. An excess from 1-3$\arcsec$
            may indicate extended emission from a starburst component. Emission from the k4 and k25
            knots in the jet is also apparent.}
\end{figure}

\section{{\it XMM-Newton} X-ray Spectra}
                                     
\subsection{RGS Spectrum}

The RGS X-ray spectrum is heavily absorbed below 1.0 keV by the large Galactic neutral absorption 
column (Fig. 5). There are strong Galactic O {\sc i}, Fe {\sc i} L,  and Ne {\sc i} K edges at 0.54, 0.71, and 0.87 keV. 
We also find Galactic O {\sc i} 1s-2p absorption at $0.5271 \pm 0.0005$ keV. This line is fit by a narrow Gaussian with an 
equivalent width of $1.1 \pm 0.3$ eV.

We initially fit the 0.3-1.8 keV RGS spectrum with a single power law absorbed by the Galactic column density, assuming
solar elemental abundances. A column density of $N_\mathrm{H}= 1.2 \pm 0.4 \times 10^{21}$ cm$^{-2}$  is measured from the 
integrated Galactic H {\sc i} 21 cm line, with an uncertainty dominated by opacity correction \citep{elw89}. The 21-cm 
measurement should be accurate to within 30\% , and includes all H {\sc i} emission within $\pm 240$ km s$^{-1}$ of the local 
standard of rest \citep{elw89, ljm86}. A survey for H {\sc i} at high velocity ($-1000<VLSR<+800$ km s$^{-1}$) yields a single cloud 
with $N_H=1.5\times 10^{18}$ cm$^{-2}$ in the direction of 3C 120 \citep{lmp02}, which is negligible. The power law model absorbed 
by the Galactic H {\sc i} column gives a very poor fit to the data, with $\chi^2_\nu=3763/943=4.0$ (Fig. 5b). 

Next we allow the Galactic O {\sc i} abundance to vary, since the O {\sc i} K edge depth otherwise limits $N_\mathrm{H}$ to 
lower values. This model gives a much better fit, with $\chi^2_\nu=1021/938=1.09$, except for some significant positive residuals 
at 1.2-1.8 keV. The best fit column densities are $N_\mathrm{H}=2.78 \pm 0.01 \times 10^{21}$ cm$^{-2}$ and 
$N_\mathrm{O}=1.55 \pm 0.03 \times 10^{18}$ cm$^{-2}$, yielding a Galactic O {\sc i} abundance of $0.66 \pm 0.01$ times the solar 
value. The $N_\mathrm{H}$ value is consistent with ROSAT measurements \citep{gsm97}. The O {\sc i} column is consistent with the 
O {\sc i} 1s-2p line EW for a Doppler parameter of $100\pm 60$ km s$^{-1}$, on the saturated part of the curve of growth. Curves of 
growth were generated with IMP, using an oscillator strength computed with HULLAC \citep{bko01,bn02}. The Galactic O {\sc i} line width 
is reasonable, considering turbulence and rotation. Because of the large uncertainty in its width, we do not rely on the 1s-2p line to measure
the O {\sc i} column, which is better measured from the K edge.

The inferred $N_\mathrm{H}$ is 2.3 times as great as that measured from Galactic H {\sc i} 21-cm emission. One possibility is
that the excess column of $N_\mathrm{H}=1.57 \pm 0.03 \times 10^{21}$ cm$^{-2}$ arises in the host galaxy or AGN itself.
Any intrinsic O {\sc i} edge should be redshifted to an energy of 0.526 keV. This edge falls near the Galactic O {\sc i} 1s-2p 
absorption line, but should be distinguished by its much broader shape (as seen for the Galactic edge). There appear to be no 
significant intrinsic absorption edges. Adding an intrinsic O {\sc i} edge to the broken power-law model, we find a $2\sigma$ upper 
limit to the column of $N_\mathrm{O,int}< 2.6 \times 10^{16}$ cm$^{-2}$, corresponding to $N_\mathrm{H,int}< 3 \times 10^{19}$ cm$^{-2}$ 
for solar abundance. An O {\sc i} abundance of $<1/50$ solar would be required if the extra hydrogen column is intrinsic to 3C 120 
rather than the Galaxy. A similar limit applies for intervening neutral hydrogen clouds, since no O {\sc i} edges are seen at redshifts
$z>0$. 

Independently increasing the Galactic He {\sc i} or C {\sc i} column over the nominal values by 
$N_\mathrm{He}=1.94^{+0.02}_{-0.05} \times 10^{20}$ cm $^{-2}$ or $N_\mathrm{C}=3.15^{+0.02}_{-0.05} \times 10^{20}$ would cause a similar amount 
of excess absorption, but yields worse fit statistics ($\chi^2_\nu=1041/938=1.11$ and $\chi^2_\nu=1108/938=1.18$, respectively).
These correspond to a Galactic He abundance of 2.6 times solar or a C abundance of 8.2 times the solar value, both of which seem
unlikely.

Alternatively, we can fit the low-energy turnover in the RGS spectrum with a broken power-law model. We fit for 
the Galactic column assuming solar O {\sc i} abundance. The best-fit model has a break at $0.561 \pm 0.005$ keV, and provides an 
excellent fit to the spectrum, with $\chi^2_\nu= 989/942 = 1.05$. Below the break energy, the spectrum is inverted with 
$\Gamma=-0.4\pm 0.1$. Above the break, the power law index is moderately soft ($\Gamma=2.00 \pm 0.01$). 
The normalization is $K = 5.36 \pm 0.04 \times 10^{-2}$ ph s$^{-1}$ cm$^{-2}$ keV$^{-1}$ at the break. 
The best fit Galactic column density is $N_\mathrm{H}=1.68 \pm 0.01 \times 10^{21}$ cm$^{-2}$, which is 40\% greater than 
that measured via H {\sc i} 21-cm emission. This corresponds to an O {\sc i} column of $1.43 \pm 0.01 \times 10^{18}$ cm$^{-2}$, 
which is close to the value we obtained above. The origin of an inverted intrinsic soft X-ray spectrum below 0.6 keV is difficult to 
understand, and it is suspicious that the spectrum breaks near the energy of the O {\sc i} edge.

\begin{figure}[t]
\plotone{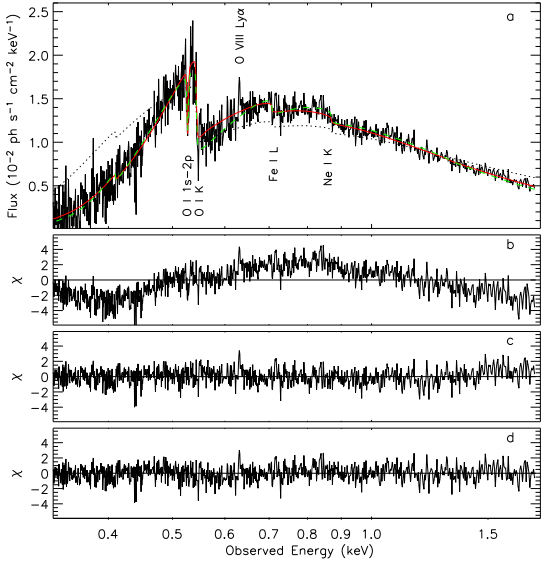}
\figcaption{(a) {\it XMM} RGS soft X-ray spectrum. Absorption line and edges from Galactic O {\sc i}, Fe {\sc i}, and Ne {\sc i}
                are indicated.  An intrinsic O {\sc viii} Ly$\alpha$ emission line appears at 0.63 keV. Dotted black line: power 
                law model with Galactic absorption. Dashed green line: Absorbed power law model with extra absorption and low O {\sc i} 
                abundance. Solid red line: broken power-law model with Galactic absorption. (b-d) Normalized residuals to these 
                3 models, respectively.}
\end{figure}

The soft X-ray continuum is not resolved into narrow emission lines at RGS resolution. This excludes a large contribution
to the soft excess from collisionally ionized or photoionized gas in the narrow- or broad-line regions. The extended emission 
at $1-3\arcsec$ from the core in the {\it Chandra} image (Fig. 4) is spatially unresolved by RGS, so any emission lines from this 
component would not be spatially broadened. In addition, it contributes only 2\% of the total soft X-ray flux. The k4 and k25 knots 
in the X-ray jet also give a negligible contribution to the spectrum.

Adding a narrow O {\sc viii} emission line significantly improves our RGS model fit by $\Delta \chi^2=29.5$ for 2 additional 
parameters. The flux in the line is $7\pm 2 \times 10^{-5}$ ph s$^{-1}$ cm$^{-2}$ and the equivalent width is $2.8 \pm 0.7$ eV.
The peak energy is $0.631 \pm 0.001$ keV, corresponding to  O {\sc viii} Ly$\alpha$ at a redshift of 0.036. 
In comparison, the optical [S {\sc ii}] emission line redshift is 0.033 (\S 5), yielding a relative velocity of 
$900 \pm 300$ km s$^{-1}$. The O {\sc viii} emission may come from hot gas which is falling toward the nucleus as the result
of a galaxy merger. Alternatively, the apparent redshift could come from absorption on the blue side of the O {\sc viii} line,
arising in an ionized outflow. Detection of additional emission lines will be required to distinguish between these two 
possibilities.

There is no other evidence of absorption from an ionized absorber. There are 
only 3 other narrow features in the residuals to the model fit which are significant at the $3\sigma$ level (Fig. 5). These 
are negative spikes at 0.438, 0.440, and 0.729 keV. With 945 bins, we expect an average of 2.5 
single bin spikes per observation to exceed $3\sigma$. We put upper limits on the column density of N {\sc vi-vii}, 
O {\sc vii-viii}, and Ne {\sc ix} of $<2\times 10^{16}-10^{17}$ cm$^{-2}$ for  $b=200$ km s$^{-1}$. For column densities 
greater than these, absorption lines would be clearly present at $>3\sigma$. The column density upper limits for inner-shell 
transitions of N {\sc i-v} and O {\sc i-iii,v-vi} are $<5\times 10^{16}-5\times10^{17}$ cm$^{-2}$ for $b=200$ km s$^{-1}$ 
(intrinsic O {\sc iv} would be blended with O {\sc i} at $z=0$). 

We find no evidence for the absorption features reported from {\it Chandra} at 0.645 keV and 0.71 keV 
\citep{mym03}. The first feature was identified with O {\sc viii} Ly$\alpha$ at $z=0.0147$
and attributed to the warm-hot intergalactic medium (IGM). The second feature is unidentified. XMM would 
have easily detected these features with their reported 2.0 eV equivalent widths if they were present 
during our observation. Fitting Gaussian absorption lines with energies and FWHMs fixed to the {\it Chandra} values, we 
find $2\sigma$ upper limits of 0.2 and 0.9 eV for the 0.645 keV and 0.71 keV features, respectively. The lack of 
an {\it XMM-Newton} detection disproves an origin in the IGM, which should not be variable in the short interval between 
the 2 observations. It also implies that either a) the features are from variable, intrinsic absorbers in 3C 120 or b) 
they are instrumental artifacts or statistical fluctuations in the {\it Chandra} spectrum.

\subsection{EPIC pn and MOS}

\subsubsection{Hard X-ray spectrum}

\begin{figure}[t]
\plotone{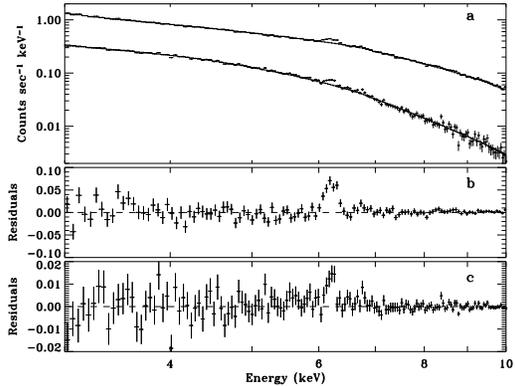}
\figcaption{(a) {\it XMM} EPIC pn (top) and MOS2 (bottom) hard X-ray spectra, Fe K$\alpha$ region. A power law plus
               neutral reflection, Fe K$\alpha$, Fe K$\beta$, and Galactic absorption are fit 
               jointly to pn and MOS2 data. (b) and (c) pn and MOS2 residuals, respectively, with the narrow line
               normalizations set to zero in the model. Marginally resolved Fe K$\alpha$ is seen at 6.21 keV, and a blend of 
               Fe K$\beta$ and Fe {\sc xxvi} Ly$\alpha$ at 6.74 keV. There is no indication of any relativistic emission lines.}
\end{figure} 

We fit the joint EPIC pn and MOS2 hard X-ray spectra (Fig. 6) with a power law from 3-10 keV. All hard X-ray  models are absorbed 
by a neutral gas column fixed at $N_\mathrm{H,Gal} = 1.7 \times 10^{21}$  cm$^{-2}$. (There is little change in the model for a
range of $N_H=1-3\times 10^{21}$ at energies $>3$ keV.) The hard power law has a photon index of $\Gamma=1.78\pm 0.01$.  The pn and MOS2 
normalizations are $K=1.36 \pm 0.02 \times 10^{-2}$ ph s$^{-1}$ cm$^{-2}$ keV$^{-1}$ and $8.6 \pm 0.1 \times 10^{-3}$ ph s$^{-1}$ cm$^{-2}$ keV$^{-1}$, 
respectively, at 1 keV. The fit is poor ($\chi^2_\nu=550/273 = 2.02$) because of residual Fe K$\alpha$ emission.

Adding a Gaussian Fe K$\alpha$ line to the model gives a greatly improved fit statistic of $\chi^2_\nu=305/270=1.13$. The line 
energy is $6.21 \pm 0.01$ keV, indicating neutral or low-ionization Fe at a redshift of $z=0.032 \pm 0.002$, consistent with the 
optical [S {\sc ii}] emission lines (\S 5). The flux in the Fe K$\alpha$ line is $3.1\pm 0.4 \times 10^{-5}$ ph s$^{-1}$ 
cm$^{-2}$, and EW$=57 \pm 7$ eV. The line is marginally resolved with an intrinsic width of $\sigma=77\pm 24 $ eV, convolved 
with the instrumental width of $\sigma_\mathrm{i}=71$ eV. Assuming a single ionization state, this corresponds to a FWHM of 
$9,000 \pm 3000$ km s$^{-1}$, indicating a possible origin in the broad-line region (BLR).

There appears to be a weaker emission line with EW$=20 \pm 7$ eV at $6.74 \pm 0.04$ keV ($6.96 \pm 0.05$ keV at $z=0.033$). 
Adding this line to the model improves the fit significantly, with $\chi^2_\nu=276/267=1.04$ ($\Delta\chi^2 = 29$ for 3
additional degrees of freedom). This may be a blend of Fe {\sc xxvi} Ly$\alpha$ (6.97 keV) and Fe {\sc i} K$\beta$ (7.06 keV). 
The ratio of the 6.75 keV to 6.21 keV line fluxes is $0.34 \pm 0.14$. This is greater than the $\sim 0.13$ expected ratio of 
neutral Fe K$\beta$ and Fe K$\alpha$ fluorescence yields \citep{km93}.

Adding Compton reflection to the model \citep{mz95} gives an equally good fit ($\chi^2_\nu= 274/266 = 1.03$). The
photon index  is not significantly altered ($\Gamma=1.82 \pm 0.04$) . The reflection fraction is weakly constrained, with 
$\Omega/2\pi =0.5 \pm 0.3$. Here we fix the high energy PL cutoff to 150 keV, the cosine of the disk 
inclination to $\cos i=0.95$ (face-on), and assume solar abundance of Fe. Adding the reflection component 
reduces the Fe K$\alpha$ line equivalent width to $49 \pm 7$ eV, and the 6.75 keV line equivalent width to
$13 \pm 5$ eV. The ratio of the 6.75 keV to 6.21 keV line fluxes is $0.27 \pm 0.11$, still marginally greater than 
the predicted  Fe K$\beta$ to Fe K$\alpha$ ratio.

We do not find any evidence of a relativistically broadened Fe K$\alpha$ line in the EPIC pn spectrum. We try fitting 
the continuum plus a Fe K$\alpha$ line with a \cite{l91} profile. A narrower line profile, such as might be found 
around a Schwarzschild black hole, would be easier to detect, so this is a conservative model. We assume a disk viewed at 
an inclination of $<20 \arcdeg$ with radial emissivity profile $R^{-q}$ , which extends from 1.24-400$R_\mathrm{G}$ around 
a near-maximally rotating Kerr black hole. A narrow Fe K$\alpha$ line and cold Compton reflection are also included in the 
model. All fits with $q=2.0-5.0$ yield zero flux for the relativistic Fe K$\alpha$ line, with an upper limit of $\sim 100-200$
eV on the equivalent width. Broader lines with $q>5.0$ are allowed, but would be indistinguishable from the underlying continuum.

We would have definitely seen a broad line as strong as the EW $=400$ eV  ($F=2.5 \times 10^{-4}$ ph s$^{-1}$ cm$^{-2}$)
line reported for the 1994 ASCA observation \citep{gsm97}. We can not exclude the possibility that the line has varied 
between 1994 and the present epoch. However, it is more likely that the Fe line width was previously overestimated 
because of the break in the underlying continuum \citep{wzs98}. Fitting the ASCA data with a power law 
that breaks at 4.0 keV yields a much smaller, narrow line with EW$=110$ eV.

\subsubsection{Soft X-ray Spectrum}

\begin{figure}[t]
\plotone{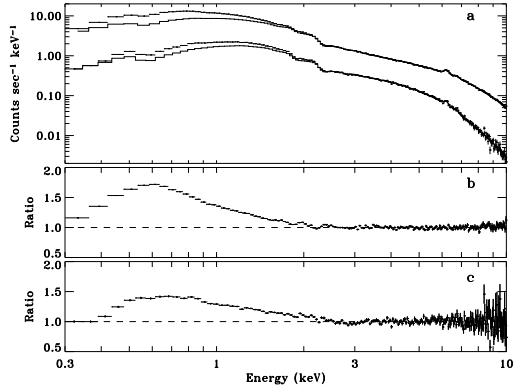}
\figcaption{(a) {\it XMM} EPIC pn (top) and MOS2 (bottom) X-ray spectra. Models are extrapolated from a 
              3-10 keV joint pn and MOS2 fit, including hard power law, neutral reflection, narrow Fe K$\alpha$ and K$\beta$, 
              and Galactic absorption. (b) and (c) Ratios of pn and MOS2 data, respectively, to extrapolated models.
              A large soft X-ray excess is evident from 0.4-2.5 keV in both data sets.}
\end{figure} 

\begin{figure}[t]
\plotone{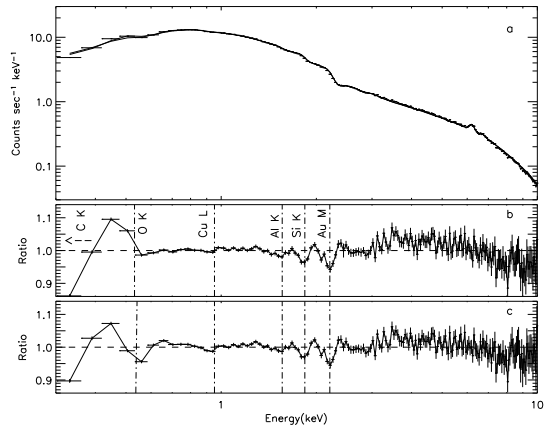}
\figcaption{(a) {\it XMM} EPIC pn 0.3-10 keV X-ray spectrum and models. (b) Ratio of data to absorbed power law model with greater than
             Galactic $N_\mathrm{H}$. c) Ratio of data to hard power-law plus soft broken power-law model. Narrow Fe K emission lines are 
             included in both models. The two models leave similar residuals, attributed primarily to instrumental C, O, Si, and Au 
             absorption edges.}
\end{figure} 

There is a large soft X-ray excess in the EPIC spectra over the extrapolated hard (3-10 keV) spectral 
models (Fig. 7). The excess is localized to the 0.4-2.5 keV band and peaks at $\sim 0.6$ keV in the pn spectrum. The MOS2 
soft excess has a smaller peak which is shifted to 0.7 keV, perhaps because of calibration differences. To further characterize 
the soft X-ray spectrum, we now concentrate on fitting the pn data, which have higher S/N and more accurate calibration than MOS2.

\begin{figure}[ht]
\plotone{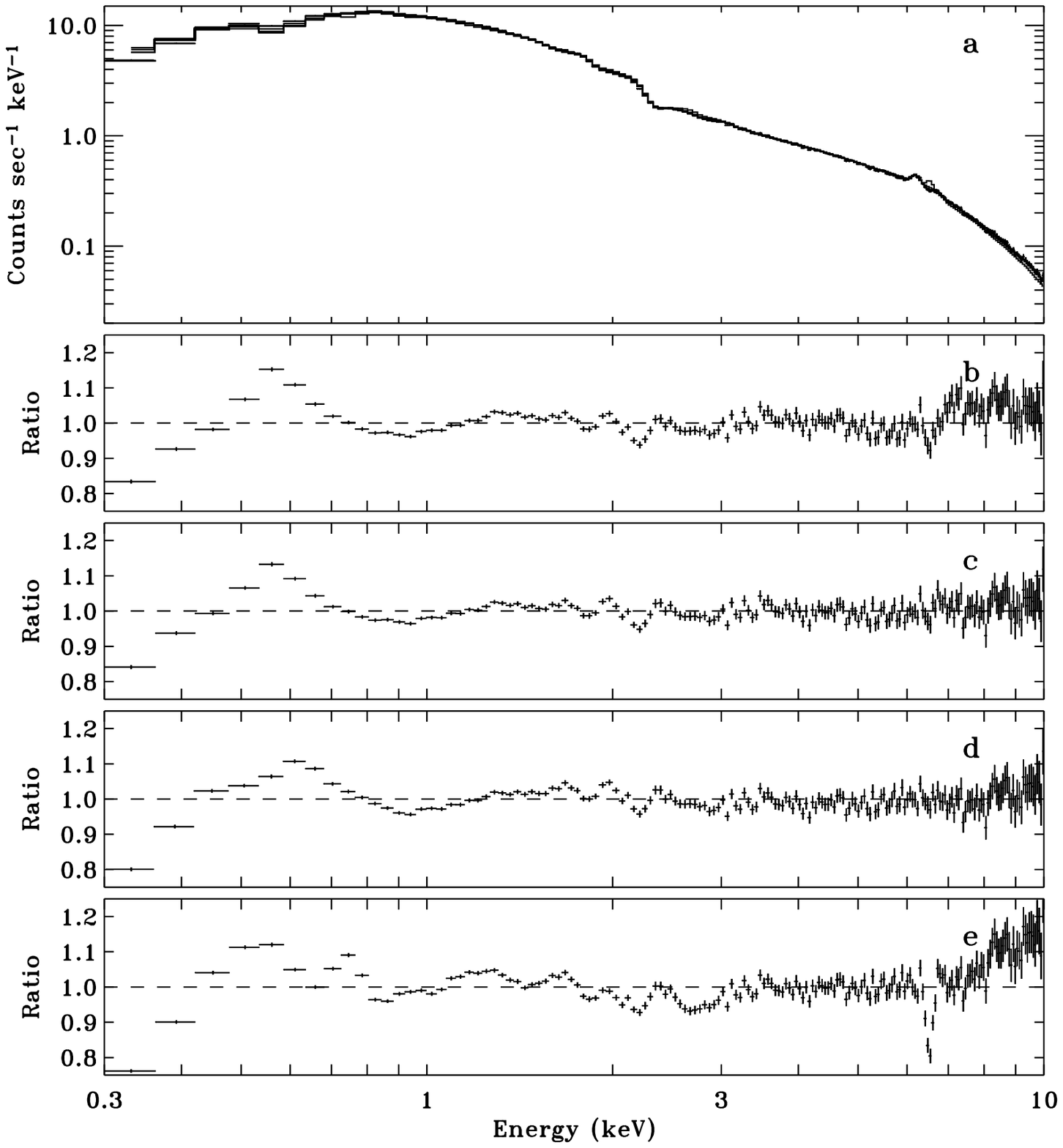}
\figcaption{(a) {\it XMM} EPIC pn X-ray data and spectral models. (b-e) Ratio of data to various models. (b) Hard power law plus neutral 
               reflection and disk blackbody soft excess, with Galactic plus intrinsic neutral absorption. (c) 
               Comptonized ({\it compTT}) model with neutral absorption. (d) Wien model for saturated Comptonization
               in a thick disk. (e) Constant density ionized disk model, plus narrow neutral Fe K$\alpha$. 
               The soft excess in this model consists of a number of blended K-shell lines from highly ionized C, N, and O 
               and Fe L-shell lines. All of these models leave unacceptably large residuals for the soft excess.}


\end{figure} 

Guided by our fits to the RGS spectrum, we parameterize the full EPIC pn continuum as a hard power law plus a
power law for the soft excess bump (Fig 8a,b). We allow the Galactic $N_H$ and O {\sc i} abundance to vary. The narrow Fe K 
emission lines are included, but no Compton reflection. This gives a formally poor fit, with ($\chi^2_\nu=1651/197=8.38$).
The spectral slopes are $\Gamma_\mathrm{soft} = 2.4 \pm 0.1$ and $\Gamma_\mathrm{hard} = 1.19\pm 0.06$. The Galactic column is
$N_\mathrm{H} = 2.2 \pm 0.1 \times 10^{21}$ cm$^{-2}$, nearly twice as great as the nominal Galactic value. 

The main residuals to the fit are caused by the C {\sc i}, O {\sc i}, and Si {\sc i} K, and Au {\sc i} M instrumental absorption 
edges. We also identify weak absorption edges from instrumental Cu {\sc i} L and Al {\sc i} K. There is curvature in
the 3-10 keV residuals, also arising from calibration inaccuracies. All of these instrumental features are seen at a similar
strength in an EPIC pn calibration spectra of 3C 273\footnote{\url{http://xmm.vilspa.esa.es/ccf/}, document XMM-SOC-CAL-TN-0018.}. 
Except for the C {\sc i} edge, the pn effective area is calibrated to better than 5\% at most energies. However, the 3C 120 observation 
has better photon statistics than this, and errors in the continuum parameters are dominated by the systematics of the instrument 
calibration. We include a 5\% uncertainty in all EPIC continuum parameters and column densities to allow for this.

Next we try a hard power law plus a broken power law for the soft excess bump (Fig. 8c). The Galactic column is fixed to 
$N_H=1.7\times 10^{21}$ cm$^{-2}$, the value obtained from the similar RGS fit. The O {\sc i} and other metal abundances are fixed to 
their solar values. This model gives a somewhat better fit, which is still formally poor ($\chi^2_\nu=5.59$). The most improvement 
occurs in the 3-10 keV region, because this model has more freedom to follow the curvature in this part of the spectrum. The 
residuals are again dominated by instrumental edges. The hard power law has $\Gamma=1.50\pm 0.08$ and dominates at $E>2.5$ keV. Below this 
energy, the soft excess follows a power law with $\Gamma=2.7 \pm 0.1$, which breaks at $0.56^{+0.01}_{-0.02}$ keV. Below the break energy, 
the soft excess has a slope of $\Gamma = -1.7^{+0.1}_{-0.7}$, which is sensitive to variations in the break energy. 

While the broken power-law soft excess provides a better fit than the power law with excess absorption, uncertainties in the 
instrument response imply that neither model is obviously preferred. Now we evaluate a number of physical models of the soft X-ray 
excess continuum which might yield a broken power-law or continuum peaked at 0.6 keV. All model fits (except the ionized disk model) 
also include a hard power law, neutral reflection, and the narrow Fe K$\alpha$ emission line. We allow for intrinsic neutral absorption 
in each, with solar abundances. 

First we fit the soft excess with a multi-color disk blackbody ({\it diskbb}) model \citep{ss73, mik84}, which would be 
appropriate for a geometrically thin, optically thick accretion disk with purely absorptive opacity (Fig. 9c). This gives a 
poor fit ($\chi^2_\nu=21$), with a residual bump peaking at 0.6 keV. The temperature at the inner radius is $T_\mathrm{in} \sim 0.2$ keV, 
too high to be attributed to a thin disk around a black hole with reverberation mass of $3\times 10^7 M_\odot$. The intrinsic neutral 
column density is $N_\mathrm{H, intr}=9 \pm 1 \times 10^{20}$ cm$^{-2}$. 

A Comptonized spectrum (XSPEC {\it compTT} model, Titarchuk 1994) also provides a poor fit to the soft excess, with  
$\chi^2_\nu=17$ (Fig. 9d). The best fit value for the Comptonizing temperature is the lowest allowed by the model (kT=2 keV), the 
temperature of the input Wien spectrum is 0.1 keV, and the Compton optical depth is $\tau_\mathrm{c} \sim 2$. It is unlikely 
a lower Comptonizing temperature would provide a reasonable fit. A problem with this model is the rather large 
temperature for the soft photon source.  If this value is fixed to something more appropriate for a big blue bump
in the range 0.01-0.05 keV, the fit becomes even worse. The intrinsic neutral column density is  
$N_\mathrm{H, intr} = 1.0 \pm 0.1 \times 10^{21}$ cm$^{-2}$, similar to the previous model.

At high accretion rates, the inner accretion flow may puff up into a thick disk or ion torus. This solution for a viscous accretion 
disk  was predicted by \cite{ss73}, in the regime where photon pressure dominates over gas pressure. For very high scattering optical 
depths and saturated Comptonization, emission from the thick disk takes on the form of a Wien bump. We fit the soft excess with a Wien 
spectrum to test this model (Fig. 9e). The best fit gives a Wien temperature of $k T_\mathrm{w}=0.18\pm 0.01$ keV ($2 \times 10^{6}$ K), 
and $\chi^2_\nu=18$. The problem with this and the previous two models is that the observed bump is cuspy (fit best by a broken power-law), 
while the models produce more gently curved bumps. Allowing a range of temperatures in the models would worsen the problem by 
broadening the predicted soft excess.

Finally, we try fitting the soft excess with a constant density ionized disk model \citep{rf93,bif01}. 
The model also includes a narrow, neutral Fe K$\alpha$ line. Adding neutral intrinsic absorption does not improve the fit, so we do not 
include it. The best fit model has $\Gamma=1.9 \pm 0.1$, a reflection fraction of $\sim 0.4$ and an ionization parameter of 
$\log \xi=\log L/nR^2\sim 2.7$. As well as providing a poor fit, with $\chi^2_\nu=35$ (Fig. 9f), the best ionized disk model produces a 
narrow 6.7 keV Fe {\sc xxv} line not seen in the data. We also tried convolving the ionized disk model with a relativistically broadened 
line profile \citep{l91}, which gives a worse fit to the EPIC pn spectrum. In comparison, \cite{brf02} find a good fit to the 0.8-10 keV 
ASCA (1994) spectrum with their ionized disk model, which includes a relativistically broadened Fe K$\alpha$ line. As we explained above, 
the appearance of a broad Fe K$\alpha$ line in the ASCA data set is most likely an artifact of the spectral break at 4 keV. The lack of any 
conspicuous broad lines in the {\it XMM-Newton} spectrum, at hard or soft energies, puts strong constraints on ionized disk models. The 
reflection fraction must be small ($<0.2$), the ionization parameter must be large ($\log \xi >4.0$), or the metal abundances must be 
sub-solar to create a featureless reflection spectrum. In any case, ionized reflection does not give the proper shape for 
the soft excess.

Lacking a viable physical model for a sharply peaked soft excess, we conclude that absorption is the most likely explanation
for the steep drop in the continuum below 0.6 keV. However, the absence of an intrinsic O {\sc i} edge is inconsistent with absorption 
from the neutral interstellar medium (ISM) of the host galaxy. A similar break is seen in {\it XMM-Newton} spectra of the BL Lac
object MS 0205.7+3509, and is attributed to a low-metallicity intervening absorber \citep{wmh04}. For a high excess H {\sc i} column 
density, we would expect to see damped Ly$\alpha$ absorption in the UV spectrum of 3C 120. There is no evidence of absorption on the blue wing of
the Ly$\alpha$ line in the {\it IUE} spectrum \citep{mcf91}, though geocoronal Ly$\alpha$ contaminates the spectrum for absorber redshifts $z<0.01$.
Another way to test the absorption hypothesis would be to search for redshifted H {\sc i} 21-cm absorption.

Next we consider the possibility that the excess absorption comes from within the jet. The spectral index below 0.6 keV is similar to that predicted 
for synchrotron self-absorption by electrons with a power-law energy distribution ($\Gamma=-1.5$, $\alpha=-2.5$, with shallower slope possible for 
multiple synchrotron components). However, the turn-over frequency seems implausibly high for synchrotron self-absorption, yielding a source size of 
only 5,000 km ($10^{-4} R_G$) and a magnetic field strength of 4 MG, assuming equipartition of electron and magnetic energies. We therefore discard 
this explanation. 

Excess absorption may also come from from matter entrained in the jet or interacting with the jet. Large turbulent velocities might smear out the 
O {\sc i} edge. In both cases, the H {\sc i} would have to be located at $r<0.5\arcsec$ in order to absorb X-ray emission from the unresolved core of
the {\it Chandra} image. However, entrainment of such a large column of H {\sc i} seems unlikely in the inner jet, which maintains an apparent speed 
of 5-6c as far as 200 mas from the core \citep{wbu01}. It is also likely that entrained matter would be rapidly ionized before reaching relativistic 
speeds. Another idea is that spallation by relativistic particles in the jet destroys the heavy metals in an interacting cloud. However, hydrogen 
would also be ionized in the process. Interestingly, \cite{gma00} attribute rapid PA rotations at $2-4$ mas from the radio core to Faraday rotation 
from an ionized cloud with a column density of $N_\mathrm{H, intr} \sim 6 \times 10^{22}$ cm$^{-2}$ which is interacting with the jet. Only $3\%$ of 
the hydrogen in the cloud need be neutral to explain the excess X-ray absorption. However, the PA rotations could also be explained by a helical 
magnetic field \citep{gma01}.

\section{Optical Spectrum}

\begin{figure}[t]
\plotone{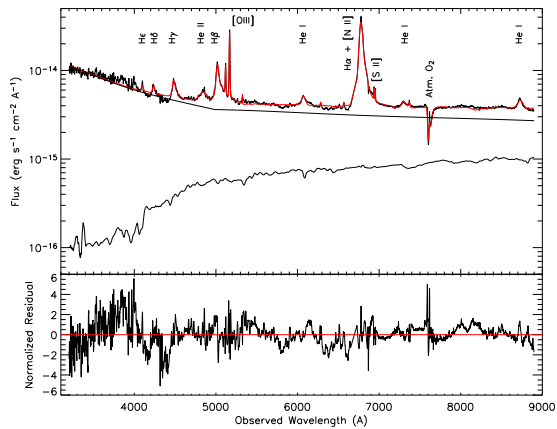}
\figcaption{Top: Keck LRIS spectrum of 3C 120. The continuum is fit with a broken power law plus an elliptical galaxy 
            template, modified by Galactic extinction. The broad and narrow emission lines are modeled with Gaussians. 
            The H-Balmer broad lines are composed of a central core and broad red wing. The absorption band at 7650 \AA\ 
            is from atmospheric O$_2$. Bottom: Residuals.}
\end{figure} 

We fit the Keck LRIS spectrum (Fig. 10) with a continuum model consisting of a broken power-law plus stellar emission, 
modified by Galactic extinction \citep{ccm89}. The LRIS spectrum also includes absorption from atmospheric oxygen and 
water bands, which is modeled using a template obtained from standard star observations. 

The extinction and reddening from dust associated with the Galactic H {\sc i} column are $A_V = 0.65$ and  E(B-V) = 0.2. Larger 
values of $A_V = 0.91$ and E(B-V) = 0.29 follow from the RGS broken power-law model with $N_\mathrm{H} = 1.7 \times 10^{21}$ cm$^{-2}$.
A reddened power law fit to the UV spectrum \citep{mcf91} gives a similar value of $A_V=0.9\pm 0.2$, and this is the
value we adopt in our models of the Keck spectrum. An intrinsic absorber with $N_\mathrm{H} = 1.6 \times 10^{21}$ cm$^{-2}$ would imply 
an additional $\Delta A_V=0.58$, if it had a Galactic dust/gas ratio. There is no evidence of this extra extinction in the optical-UV spectrum, 
which agrees with the low O {\sc i} abundance from the X-ray spectrum.

We use the elliptical galaxy NGC 1316 to create a template for starlight emission from the host galaxy. The galaxy 
fraction increases from 7\% at 5000 \AA\ to 28\% at 8900 \AA. We can not make an accurate measurement of the host galaxy
redshift because the galaxy fraction is small and the absorption lines are confused with emission lines from the AGN. 

The underlying (extinction-corrected) optical continuum is approximated by a broken power-law with indices of $\alpha_1 = -1.71\pm 0.01$ 
and $\alpha_2 = -0.50\pm0.02$, respectively, below and above a break at $\lambda=4987^{+3}_{-8}$ \AA. It is likely that a pseudo-continuum 
of broad Fe emission lines and Balmer continuum contribute significantly below 4000 \AA. The power-law normalization is 
$A_{\lambda}=1.1\pm 0.1 \times 10^{-14}$ erg s$^{-1}$ cm$^{-2}$ \AA $^{-1}$ at the break. 

The optical spectrum contains a number of broad permitted lines and narrow forbidden lines. The best-fit emission line 
velocities, widths, and extinction-corrected fluxes are given in Table 2. Emission lines from the same ion are constrained 
to have the same redshift and Doppler width. The [S {\sc ii}] lines are the narrowest in the spectrum (FWHM$=330\pm 40$ km s$^{-1}$),
so we use their redshift of $z_n=0.03300\pm 0.00006$ to compute relative velocities for the other lines. The narrow 
$[$O {\sc iii}$]$ emission  lines have a greater width of FWHM $= 560\pm 20 $ km s$^{-1}$ and are blue-shifted by 
$-280\pm 7$ km s$^{-1}$ with respect to [S {\sc ii}]. Weak narrow [Fe {\sc vii}], [Ar {\sc iii}], and [O{\sc i}] are also 
visible. We fail to detect [O {\sc ii}] $\lambda$3727 which is in a noisy part of the spectrum, or [N {\sc ii}] 
$\lambda \lambda$6548,6583 which are blended with broad H$\alpha$.

The broad He {\sc i} emission lines are redshifted  by $200\pm 60$ km s$^{-1}$, and have widths of FWHM$=2700\pm 200$ km s$^{-1}$ 
(Table 2). The He {\sc ii} line has a similar redshift, but is considerably broader, with  FWHM$=6300 \pm 400$ km s$^{-1}$.
The broad H-Balmer emission lines have asymmetric profiles, with a broad red tail. A similar red asymmetry is seen in the 
UV broad line profiles of H {\sc i} Ly$\alpha$, and C {\sc iv} $\lambda$1549 \citep{mcf91}. We have decomposed the broad 
H-Balmer lines into three Gaussian components, one narrow, and two broad. The narrow (FWHM$=1970\pm 30$ km s$^{-1}$) component 
has a small blueshift of $-110\pm 20$ km s$^{-1}$. The broad components have large redshifts of $1070\pm90$, and $5600 \pm 300$ 
km s$^{-1}$, respectively. The red asymmetry may be attributed to optical depth effects if the Balmer lines are emitted in a disk wind.
The very broad component (FWHM$=13,700\pm 600$ km s$^{-1}$) is most prominent in the H$\beta$ line, but this may be contaminated
by emission from broad Fe II. 

The total (extinction-corrected) H$\beta$ flux is $1.85\pm 0.05\times 10^{-12}$ erg s$^{-1}$ cm$^{-2}$. The 
reddening-corrected Balmer series lines are in the ratios H$\alpha$ : H$\beta$ : H$\gamma = 2.62 : 1.00 : 0.27$. 
The Balmer decrement is significantly less than the value of 3.1 predicted for for Case B recombination, which has 
H$\alpha$ : H$\beta$ : H$\gamma =  3.1 : 1.0 : 0.46$ \citep{om75}. It is also smaller than the average Balmer decrement 
of 3.5 for the \cite{pmb82} sample of Seyferts and quasars \citep{wdf88}. This may indicate that we have over-corrected 
for reddening. However, if we remove the broadest Balmer line component, which is potentially contaminated
by broad Fe II,  we find H$\alpha$ : H$\beta$ : H$\gamma = 4.12 : 1.00 : 0.46$. The Balmer decrement therefore depends 
sensitively on how the line is decomposed. 

The Balmer H$\beta$ equivalent width of $100\pm 4$ \AA\ is greater than the mean AGN value of $\sim 90$ \AA\ \citep{dhs02}. 
(Note that H$\beta$ EW is fairly independent of continuum luminosity.)  
This indicates a negligible amount of beamed continuum at 5000 \AA. This is corroborated by the relatively low optical 
continuum polarization ($P=0.9-1.2\%$) which is unlikely to have a synchrotron origin since it shares the same position
angle (PA) as the emission lines  \citep{rss83, ant84}. The polarization drops to 0.3\% in the H band \citep{bhb90}, 
arguing against a near-IR synchrotron component. The contribution to the optical spectrum from the extended jet is 
negligible, since its extinction-corrected B-band flux density of $14 \mu$Jy \citep{hvs95} is less than 0.1\% of the flux 
density from the core. We conclude that highly beamed continuum from a relativistic jet does not greatly compromise our 
estimate of the isotropic AGN luminosity and accretion rate. 

\section{Discussion}

\subsection{Spectral Energy Distribution}

The spectral energy distribution (SED) of 3C 120 can be used to constrain the properties of the accretion flow and
the jet. The optical through soft X-ray portion of the SED is affected by extinction and reddening from a large Galactic 
column. We correct the Keck spectrum and OM photometry for this using the extinction law of \cite{ccm89}, with $A_V=0.9$ mag 
and $R_\mathrm{V}=3.1$. The X-ray spectrum is corrected for a Galactic absorption column of $N_\mathrm{H}=1.7 \times 10^{21}$ cm$^{-2}$. 
Our data points for the SED are compiled in Table 1 and plotted in  Figure 11a. Except for the high 250 GHz flux, our data are in good 
agreement with the archival SED (Fig. 11b) from the NASA Extragalactic Database (NED) and the IR to X-ray SED measured by \cite{mcf91}.

The extinction and absorption-corrected luminosity from optical through X-rays ($9\times 10^{14}$-$3\times 10^{18}$ Hz) is 
$L_\mathrm{ox}\sim 1.3\times 10^{45}$ erg s$^{-1}$ (assuming isotropic emission and a Hubble constant of $H_0=70$ km s$^{-1}$ 
Mpc$^{-1}$). We have interpolated a power-law with $\alpha=2.2$ between the UVW1 and 0.3 keV flux points to estimate the UV 
luminosity. In comparison, the Eddington luminosity for a $3\times 10^7 M_\odot$ black hole is $4\times 10^{45}$ erg s$^{-1}$, 
yielding an Eddington ratio of $\sim 0.3$. The accretion flow is therefore radiatively efficient, contrary to the idea that 
3C 120 contains an advection-dominated accretion flow \citep{exm00}.

\begin{figure}[t]
\plotone{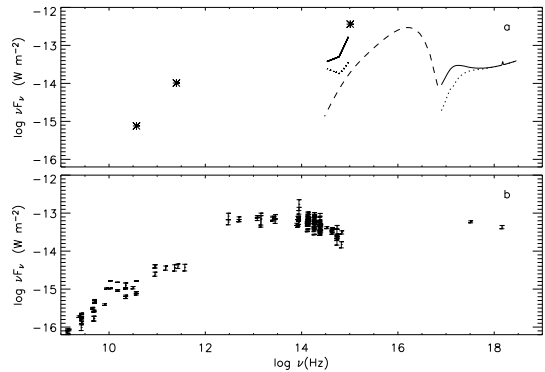}
\figcaption{3C 120 spectral energy distribution. (a) August 2003 epoch (this work). The galaxy-subtracted optical continuum
            and best EPIC pn X-ray model (solid lines) have been corrected for Galactic extinction and photoelectric
            absorption. The dotted line is the uncorrected continuum.  Radio and extinction-corrected UVW1 fluxes are 
            plotted as asterisks. For comparison, we plot a multi-color disk blackbody spectrum (dashed line) for a black hole mass of 
            $3\times10^7 M_\odot$ and luminosity of $L=0.3 L_E $. 
            (b) Flux points from NASA Extragalactic Database (1967-2003).}
\end{figure} 

The optical-FUV region is crucial for studying any thermal accretion disk contribution to the SED, typically manifest 
as the so-called big blue bump (BBB). For comparison to the observed SED, we plot a multi-color disk blackbody spectrum 
for a $3 \times 10^7 M_\odot$ black hole radiating at 30\% of the Eddington luminosity. The disk extends from 
$R= 1.8-400 GM/c^2$, and we neglect relativistic effects and Comptonization. This model has an optical spectral index of 
$\alpha=-0.3$, which does not match the observed (de-reddened) optical index of $\alpha=-0.50\pm 0.02$. In addition, the observed 
flux is a factor of $\sim 7$ greater at 5000 \AA\ than the model prediction. The excess can not be accounted for by 
emission from the broad-line region. A simple multi-color disk model gives a poor description of the observed SED. 

The ratio of 0.3-10 keV X-ray to optical luminosity is $0.18$, so a significant fraction of the accretion energy may be dissipated in a 
hot corona. The hard X-ray power law is consistent with unsaturated thermal Comptonization in a hot plasma with a Compton y-parameter of 
1.3. The soft X-ray excess contributes 37\% of the total 0.3-10 keV X-ray luminosity, so it is also energetically important. A soft
excess is not predicted in simple corona models, and requires an additional emission or scattering region. A second, cooler 
Compton-scattering region may produce the soft excess \citep{mbz98}. Perhaps a disk corona produces the soft excess while the hard power
law comes from the base of the jet. Another possibility is that the soft excess arises from turbulent Comptonization in the disk \citep{sdb04}. 
However, none of these models predict a break in the spectrum below 0.6 keV.

The reddening-corrected optical to X-ray spectral index (measured between 2900 \AA\ and 2 keV) is $\alpha_\mathrm{OX}= -1.4 \pm 0.1$, 
which is typical for an AGN or quasar \citep{mw89}. There is no indication that the radio-loudness of the source has any 
effect on the relative strengths of the UV and X-ray emission. It would be valuable to determine if $\alpha_\mathrm{OX}$ 
changes during radio component ejection events \citep{mjg02}. This would test if there is a connection between these events
and the state of the inner accretion disk. 

In principle, the X-ray continuum could be enhanced by beamed emission from the relativistic jet. The radio-X-ray index is 
$\alpha_\mathrm{rx}=0.79\pm 0.01$ (37 GHz-2 keV), similar to the index expected for optically thin X-ray synchrotron emission. This also 
matches the observed hard X-ray spectral index, consistent with a synchrotron spectrum extending directly from radio through hard X-rays. 
However, this appears to be a coincidence since the strong Fe K$\alpha$ emission line rules out a highly beamed X-ray continuum (\S 6.2). 
The 250 GHz to X-ray index of $\alpha_\mathrm{mx}=-0.94\pm 0.01$ is not steep enough for synchrotron emission from the 250 GHz radio component 
to account for much of the soft X-ray excess. We note that while there does not appear to be a large amount of beamed X-ray emission from the 
jet, it is entirely possible that unbeamed X-rays are produced in a sub-relativistic jet base. Or equivalently, the hard X-ray emitting corona 
may be the jet base. This is supported by the strong correlation observed between X-ray and (beaming-corrected) radio core luminosity in AGN
\citep{mhd03,fkm04}. 

\subsection{Fe K$\alpha$ Emission Line}

The Fe K$\alpha$ emission line can be used to give an indication of the beamed continuum fraction. 
For an isotropic continuum source, it gives an estimate of the covering fraction of the emission line region. The velocity width
of the line can be used to estimate the radial distance of the emission line region from the nucleus given the black hole mass.

The strength of the Fe K$\alpha$ line (EW$=57 \pm 7$ eV) is evidence that the hard X-ray continuum originates primarily from an 
unbeamed (or weakly beamed) source such as an accretion disk or base of the jet rather than the superluminal jet. If all of this
continuum were highly beamed within a $20\arcdeg$ cone, then the geometry would be unfavorable for reflection. For neutral 
fluorescence, the line equivalent width scales with the reflection fraction roughly as EW$=(\Omega/2\pi) 150$ eV \citep{zg01,gf91}
(for $\Gamma=1.8$). The observed Fe K$\alpha$ equivalent width yields $\Omega/2\pi = 0.38 \pm 0.05$, consistent with the value 
derived for Compton reflection in the hard X-ray spectrum \citep{zg01}.  This implies a beam opening half-angle for the hard 
X-ray continuum of $>50\arcdeg$. 

If the hard X-rays are unbeamed and reprocessed by a molecular torus, then the Fe K$\alpha$ line in 3C 120 implies
a torus opening half-angle of $\sim 70\arcdeg$, in the limit of large torus column density. In comparison, the mean narrow 
Fe K$\alpha$ equivalent width for Seyfert 1s observed with ASCA \citep{n97} is 100 eV. This implies a larger mean covering fraction of 
0.63 and a smaller mean torus opening half-angle of $51\arcdeg$ for Seyferts. If the intrinsic Fe K$\alpha$ line EW of 3C 120 is truly 
smaller than that of the average Seyfert 1, then the maximum fraction of beamed hard X-ray continuum is 40\%.

The Fe K$\alpha$ line is marginally resolved by EPIC pn, with FWHM$=9,000\pm3,000$ km s$^{-1}$. This indicates that
at least part of the line may come from the outer accretion disk or BLR. Assuming that the emission region is virialized 
gives a distance of $\sim 4000 R_G$ from the central black hole. In that case, the light crossing time is 
15 days, so we do not expect significant Fe K$\alpha$ variability during the {\it XMM-Newton} observation. However, this is subject to considerable
uncertainties in the geometry and dynamics of the Fe K$\alpha$ emission region.  The broadest lines in the optical spectrum have comparable 
FWHM$=6000-13,000$ km s$^{-1}$. It is an intriguing possibility that Fe K$\alpha$ and the optical broad lines come from the same region. 
In that case, the torus covering fraction arguments above apply to the BLR instead.

Line blending and other broadening mechanisms are unlikely to account for the Fe K$\alpha$ line width.
The separation between the lines in the Fe {\sc i} K$\alpha$ doublet is 13 eV \citep{dpo95}, much narrower than the observed 
width. The expected energies of the Fe {\sc xviii} and  Fe {\sc i} K$\alpha$ lines are separated by 50 eV, again less than 
the observed line width. The Fe K$\alpha$ line center is within 10 eV (at 90\% confidence) of the expected energy of Fe {\sc i} 
at $z=0.033$, arguing against a large contribution from ionization states greater than Fe {\sc xv}. The natural and
thermal line widths are also negligible in a low-ionization plasma \citep{ygn01}. The gravitational redshift at $4000 R_G$
is $\sim 1.5$ eV, which is small.

The non-detection of relativistic Fe K$\alpha$ in the EPIC spectra does not constrain neutral accretion disk models 
particularly well. The 100-200 eV upper limit for the line EW corresponds to a cold reflection fraction of $\Omega/2\pi < 0.7-1.3$. 
This still allows a simple model where the hard X-rays are produced entirely in a plane-parallel, isotropically emitting corona over a 
cold disk with solar iron abundance. In comparison, the relativistic Fe K$\alpha$ line in MCG-6-30-15 has an equivalent width of 
300-400 eV, and appears to come from an {\it ionized} disk \citep{wrb01}. We would have detected a similar line in 3C 120 with great 
confidence. However, such lines appear in only a small number of Seyfert 1s observed with {\it XMM-Newton}, and may require 
special conditions to form. It is not clear that the absence of relativistic Fe K$\alpha$ in 3C 120 is related to 
its status as a radio-loud AGN.


\subsection{X-ray Variability}

\begin{figure}[t]
\plotone{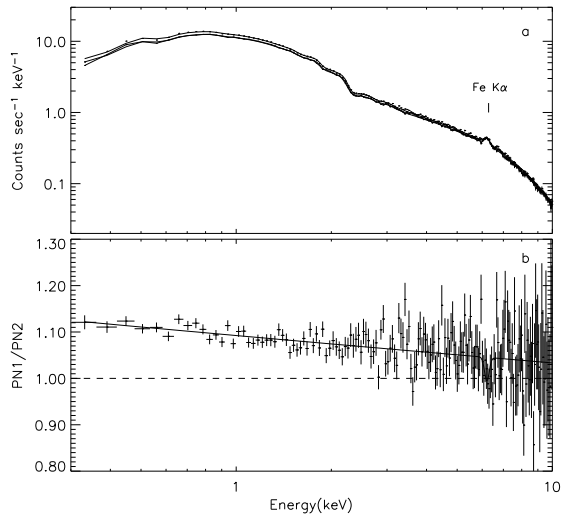}
\figcaption{{\it XMM} EPIC pn X-ray spectral variability. (a) Counts spectra for 1st and 2nd halves of the observation.
            (b) Ratio of first 60 ks spectrum to second 60 ks spectrum. The spectrum appears to
              pivot at $E>10$ keV. Fe K$\alpha$ (at 6.2 keV) is constant to within the uncertainties of the measurement.}
\end{figure} 

\begin{figure}[t]
\plotone{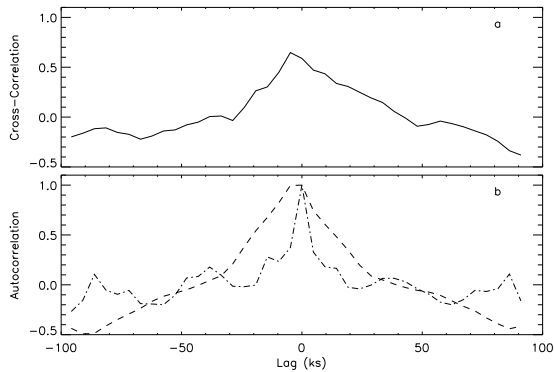}
\figcaption{(a) Cross-correlation function (CCF) between soft (0.3-2 keV) X-ray and UVW1 light curves. The strong
             peak indicates a highly significant correlation between X-ray and UV flux.
            (b) Autocorrelation functions for soft X-ray (dash), and UVW1 (dot-dash). The CCF is
             a convolution of these with the X-UV response function.}
\end{figure} 

The X-ray flux of 3C 120 decreases in a fairly steady manner during the {\it XMM-Newton} observation
(Fig. 1a,b), accompanied by a gradual increase in the hardness ratio (Fig. 1c). The hard (2-10 keV) count rate
drops by 13\%, while the soft (0.3-2 keV) count rate drops by 18\% in 120 ks. The net result is a hardening of
the spectrum and an anti-correlation between X-ray flux and spectral hardness. The hard X-ray flux doubling time 
is roughly $10^6$ sec, corresponding to a light-crossing time of $<12$ days ($r<3400 R_G$).

The correlation between X-ray spectral 
index and soft X-ray flux was discovered in a 1979-1981 series of {\it Einstein} observations, and attributed to a 
variable, steep synchrotron component in the soft X-ray spectrum \citep{h85}. Further {\it EXOSAT} observations show that 
the X-ray spectrum sometimes pivots at an energy of 2 keV \citep{mcf91}, arguing against the synchrotron interpretation. 
The correlation persists at shorter timescales, as seen in a 200 ks {\it Beppo-SAX} observation by \cite{zg01}, who
find a pivot energy of  $\sim 5$ keV. These authors suggest that the Comptonized X-ray continuum responds to variations in 
the UV seed photon flux. An increase in the seed photon flux cools the Comptonizing atmosphere, resulting in a steeper X-ray 
spectrum. 

To better determine the form of spectral variability, we split the EPIC pn spectrum into two 60-ks halves (Fig. 12). 
The ratio of the 2 spectra (Fig. 12b) demonstrates that the second spectrum is harder than the first across the entire
0.3-10 keV band. There is no break in the ratio at either of the X-ray spectral breaks. This implies a close 
connection between the hard and soft X-ray variability. The ratio is fit well by a power law ($\chi^2_\nu=202/179=1.13$), with a 
slope of $\Gamma=-0.024\pm 0.001$ and a pivot energy of $39 \pm 3$ keV (extrapolated beyond the EPIC spectral range). This is 
much greater than the pivot energies found by \cite{mcf91} and \cite{zg01}, and consistent with the $>6$ keV pivot found by 
\cite{h85}. Most Seyfert galaxies which display pivoting have pivot energies of $>10$ keV \citep{zlg03}. The occasionally low 
pivot energy in 3C 120 is unusual, but can be explained if the corona luminosity and optical depth stay constant while the seed 
photon flux changes \citep{zg01}. 

The Fe K$\alpha$ line is constant within the uncertainties of our measurement, resulting in a (noisy) dip in the ratio spectrum at 
6.2 keV (Fig. 12b). Adding a Gaussian to model the dip improves the fit at the 95\% level ($\Delta \chi^2 =4.2$ for 1 additional 
degree of freedom), illustrating a small ($5 \pm 4$ eV) rise in the Fe K$\alpha$ equivalent width. The constant Fe K$\alpha$
flux is consistent with an origin in the BLR at $r \sim 8$ light-days, which is not expected to vary much during the {\it XMM-Newton}
observation.

\subsection{UV Variability}

The UVW1 count rate is variable, falling by 2\% along with the X-ray flux (Fig. 1d). This variability is shown to be highly
significant by a $\chi^2$ test, which gives $\chi^2_\nu=132/25 = 5.3$ (with a vanishingly small probability).  It is unlikely 
that Mg II emission from the BLR varies much over the 1.5 day duration of the OM observation. A constant, $<7\%$ contribution 
of Mg {\sc ii} to the total flux reduces the observed variability amplitude in the UVW1 band by only 0.1\%.

We compute the cross-correlation function (CCF) between the 0.3-2 keV X-ray and UVW1 light curves (Fig. 13), binning the X-ray 
curve to match the UVW1 sampling. We use interpolation to compute the CCF, according to the method of \cite{wp94}. There is a 
strong peak in the CCF with a peak amplitude of 0.65 at zero lag. To determine the significance of the correlation, we simulate 
1000 independent X-ray light curves and cross-correlate them with the UVW1 light curve. The duration, sampling, and variance are 
matched to the observed X-ray curve. We assume a red noise prescription for the power spectral density (PSD) of the X-ray light 
curve, which follows a $\nu^{-2}$ power law. Our Monte Carlo simulations show that peaks as strong as the observed one occur by 
chance only $1.4 \pm 0.4 \%$ of the time. We therefore conclude that the soft X-ray and UV light curves are positively correlated. 

However, the correlation doesn't necessarily imply the UV and X-ray fluxes are causally connected.
If we divide out the overall slopes of the two light curves and repeat the cross correlation, we find no significant correlation 
(peak amplitude of 0.31). Consequently, it is not possible to determine a reliable lag between the UV and X-ray light curves. A 
longer data stream with correlated ups and downs would be necessary to determine a lag.
 
The UV-X correlation could in principle be caused by X-ray heating of the UV emitting region.  The 2\% magnitude of the 
UVW1 variation is consistent with $L_\mathrm{x}/L_\mathrm{ox}=0.18$ from the SED times the 15\% X-ray variability, if
the UV emitting region intercepts all of the X-ray flux. If we assume an accretion disk which  radiates locally like a 
black body, then the 2900 \AA\ emissivity peaks at $200 R_G$ for a $3\times 10^7 M_\odot$ black hole accreting at 30\% of the 
Eddington rate. We would expect the UV to follow the X-rays with a perceptible delay ($\sim 30$ ks light-travel time) 
if the X-ray source is located near the center of a disk. It will be important to measure the lag to test this
hypothesis. Such X-ray reprocessing may have been detected in the UVW1 light curve of NGC 4051 \citep{mmp02}, where the UV 
variations lag the X-rays by $\sim 20$ ks. Alternatively, the X-rays could be produced directly above the UV emission region,
yielding no perceptible lag.

Another possibility is that we have observed the variability signature of Compton 
scattering of UV photons into the X-ray band. The drop in UVW1 (2900 \AA) photon flux seems insufficient to drive the greater 
drop in X-ray photon flux. However, there may be greater variability at shorter UV wavelengths. We expect the X-rays to lag the UV
if they are produced by Comptonization of UV photons. The size of the lag depends on the geometry of the X-ray and UV emission
regions. If the X-ray emission comes from a coronal layer above the accretion disk, the delay will be small. There also may be 
positive feedback, with the Comptonized photons scattered back to the disk providing an extra source of heating. 

\subsection{Radio Variability}

The X-ray variability should be considered in the context of a strong radio flare (Fig. 2). The {\it XMM-Newton} observation
took place when the 250 GHz emission from the source was at an all-time high. The combined Mets\"{a}hovi and SEST flux 
points for 18 August 2003 indicate an inverted radio spectrum ($F_\nu \sim \nu^{0.3}$). We may have witnessed a high-frequency 
peaking radio flare, associated with a new radio component ejection. We believe that the flare in these cases starts at very 
high (i.e., IR) frequencies. Then the shock grows in strength, and its spectrum moves in ($S$, $\nu$) space so that the 
turn-over frequency decreases and the shocked region becomes visible at 250 GHz, then at lower radio frequencies. All indications, 
from emission line equivalent widths and continuum variability, are that the flare and radio jet do not contribute a significant 
flux in the optical or X-ray bands.

The 37 GHz outburst in 2003 had a magnitude of 2 Jy and a 17 day exponential rise time. The corresponding variability brightness 
temperature \citep{lv99} is $T_\mathrm{b,var} = 1.5 \times 10^{12}$ K. Assuming that the flaring component has an equipartition 
temperature of $5\times 10^{10}$ K \citep{r94}, the variability Doppler factor is $D_\mathrm{var}=3.1$, similar to the Doppler factor
derived from high-frequency VLBI measurements at earlier epochs \citep{gma00}. Looking at the historical 37 GHz flux curves of 3C 120 
\citep{ttm98} we see that the timescales of the flares are much slower than for the 2003.7 event, with an average rise time of $\sim 70$ 
days. The corresponding average $D_\mathrm{var}$ is small ($\sim 1$), with variability brightness temperatures close to equipartition.
The relatively low $D_\mathrm{var}$ values for 3C 120 are consistent with its classification as a radio galaxy. In comparison, BL Lac objects 
and quasars have median values of $D_\mathrm{var}=5-11$ \citep{lv99}.

The timescale for the fastest flare (at 22 and 37 GHz) in each of the 85 sources in the sample of \cite{vlt99} 
and \cite{lv99} has a median of 60 days for high polarization quasars (HPQs), 90 days for BL Lac objects and $\sim 120$
days for low polarization quasars (LPQs) and radio galaxies. Only 3 out of 85 have a timescale less than 20 days 
and an additional 5 have a timescale less than 30 days. These are all HPQs and BL Lac objects. It's even rarer for a 
source to change its behavior so suddenly and dramatically--there seem to be no other cases at 37 GHz where a source 
exhibiting average timescale flares for many years would suddenly flare up extremely fast (A. L\"{a}hteenm\"{a}ki, 
private communication). The 2003.7 radio flare of 3C 120 is very unusual in this respect. The short timescale of the 
flare indicates an origin in a very compact region ($<4\times 10^{16}$ cm 
$= 1\times 10^{-2}$ pc $= 9 \times 10^3 R_G$).

The 2003.7 radio flare is comparable in flux to the biggest flare observed by \cite{mjg02}. That 1998 
flare was preceded by an X-ray dip. New {\it RXTE} observations show that there was an X-ray dip at 2003.5, 47 days prior to
the {\it XMM-Newton} observation, and 54 days prior to the start of the 37 GHz flare \citep{mja04}. 
The mean delay between the X-ray minima and the ejection of VLBI components is $37 \pm 11$ days, for the first 4 dips 
observed. If the 2003.7 flare follows a similar component ejection (which remains to be seen), then we expect that
the ejection date preceded the 37 GHz flare by about 17 days. 

{\it RXTE} monitoring data is available for the years 1997-2000, and 2002-2004 \citep{mja04}. The Galactic absorption-corrected
{\it RXTE} 2-10 keV flux for 2003.65 was $5.8\pm 0.2\times 10^{-11}$ erg s$^{-1}$ cm$^{-2}$, compared to a mean of 
$4.7\pm 0.2 \times 10^{-11}$ erg s$^{-1}$ cm$^{-2}$ from EPIC PN on the following day. The {\it RXTE} flux was 20\% greater, 
consistent with the decreasing flux during the {\it XMM-Newton} observation. The {\it RXTE} 2.4-20 keV spectral index was 
$\alpha=0.69\pm 0.05$, 0.1 dex harder than the EPIC PN 2-10 keV spectral index. The X-ray spectral index during the {\it XMM-Newton} 
observation is relatively hard, but not as hard as the $\alpha=0.4$ value for the 2003.5 X-ray minimum observed by \cite{mja04}. In any case, it 
appears that 3C 120 was in a fairly normal X-ray state, post dip and perhaps coincident with an ongoing 250 GHz flare.

\section{Conclusions}

The X-ray spectrum of 3C 120 shows a relatively strong Fe K$\alpha$ emission line at 6.21 keV, and a weaker line
at 6.74 keV which may be a blend of  Fe K$\beta$ and Fe {\sc xxvi} Ly$\alpha$. The Fe K$\alpha$ line energy 
is consistent with low ionization Fe {\sc i-xv} at a redshift of 0.033. This line is marginally broadened,
with FWHM$=9,000\pm 3,000$ km s$^{-1}$, similar to the width of the optical broad lines. The equivalent width of $57\pm 7$ eV
implies a cold gas covering fraction of $\sim 0.4$, about 40\% less than in the average Seyfert galaxy. This is consistent 
with the covering fraction of the neutral Compton reflector, which likely arises in the same location, far from the central 
black hole. The strength of the Fe K$\alpha$ line excludes a strong beamed component from the jet in hard X-rays. 

There is no indication of the relativistically broadened Fe K$\alpha$ emission line found in {\it ASCA} data, nor 
do we find any relativistically broadened soft X-ray lines. As suggested by previous investigators, it is likely that the 
broad wing on the Fe K$\alpha$ line was an artifact of fitting a single power-law to a continuum which is better 
described by a broken power-law. We estimate an upper limit of 100-200 eV for any relativistic emission from the
inner accretion disk. This does not put strong constraints on accretion disk models.

The lack of relativistic emission lines is not unique to 3C 120 or radio galaxies in general. Therefore, we can not make any 
firm conclusions from this concerning the inner accretion disk structure of radio-loud vs. radio-quiet AGN. In fact, it appears that 
strong relativistic emission lines may be restricted to a subset of the class of radio-quiet narrow line Seyfert 1 galaxies \citep{br01,omp04}. 
These are thought to contain low-mass, near-maximally spinning black holes, accreting at close to the Eddington rate. They also often 
have winds with high outflow rates. Interestingly, this indicates that the condition of high black hole spin plus high accretion rate is 
adverse to radio jet formation. 3C 120 appears to accrete at $\sim 0.3$ times the Eddington rate, so the accretion rate can not be 
considered particularly low or high. However, this is fairly efficient, casting doubt on models with a truncated disk and inner 
advectively dominated accretion flow. 

During the {\it XMM-Newton} observation, the X-ray flux dropped by 15\% and hardened. This is consistent with previous 
behavior, and indicates pivoting with a pivot energy of $E \sim 40$ keV. At the same time, the UV flux dropped by 2\%, in 
a strongly correlated fashion. This is the first time such a strong correlation has been found between UV and X-ray emission 
from 3C 120 at the 10-100 ks timescale. Further observations are necessary to determine if there is a lag between the UV and X-ray 
variations. The sign of the lag may help to distinguish between X-ray reprocessing in an accretion disk and Compton 
up-scattering of UV photons. The X-ray spectral hardening, correlated with the drop in UV flux, leads us to favor Compton 
scattering.

There is a strong soft X-ray excess which comprises nearly 40\% of the X-ray flux and may indicate a second source of 
Comptonized emission. There is no indication that this emission comes from the relativistic jet. Extended thermal
emission contributes $<2\%$ of the total 2-12 keV X-ray flux in an archival {\it Chandra} HETGS image, and may arise
from starburst regions in the disturbed spiral host galaxy. The soft excess varies together with the hard X-ray power law, 
indicating a close connection between the two components. We suggest that the soft excess comes from a corona above the disk while 
the hard power-law comes from the sub-relativistic base of the jet.

The X-ray spectrum appears to be absorbed below 0.6 keV by an intrinsic or intervening absorber with a column density of 
$N_\mathrm{H}=1.57 \pm 0.03 \times 10^{21}$ cm$^{-2}$. The lack of an O {\sc i} K edge from this absorber implies an oxygen abundance 
$<1/50$ of the solar value. Such a high-column, low-metallicity absorber could be an intervening proto-galaxy. Alternatively, the 
absorber may come from material interacting with the jet. H {\i} 21 cm absorption studies may help to resolve this issue.

There are no absorption lines or edges from an ionized (warm) absorber.
The lack of a warm X-ray absorber is consistent with a line of sight that passes close to the axis of the accretion
disk and jet. The jet may clear a channel which is free of warm ionized gas. We detect a narrow O {\sc viii}
Ly$\alpha$ emission line in the soft X-ray spectrum, which may be from an ionized absorber out of the line of sight.
Our observations do not support a previous claim of absorption from the ionized intergalactic medium. We detect no such lines in 
the highest S/N X-ray spectrum to date.

We caught 3C 120 during a 37 GHz radio flare which started just 8 days after the {\it XMM-Newton} observation.
The flux at 250 GHz was higher than it has ever been observed before. The flare was also unusual in having a very short
rise timescale of 17 days, indicating emission from a very compact region. The radio flare was preceded 54 days by an 
X-ray dip observed with {\it RXTE}, and may indicate a new VLBI radio component. There is no obvious connection between
the radio and X-ray emission over the 1.2 day timescale of the {\it XMM-Newton} observation. However, the X-ray dips observed
with {\it RXTE} have a short duration of $\sim 2-6$ days, after which 3C 120 returns to a normal X-ray state. An
observation of 3C 120 during one of these dips would more closely probe the connection between the X-ray spectrum and ejection
of VLBI radio components. This will require vigilant monitoring and a rapid response time of $\sim 1$ day.

\acknowledgements

We thank Alan Marscher for providing {\it RXTE} data in advance of publication, and Y. Chin for assisting with SEST
observations. P. M. Ogle is supported by a National Research Council research associateship. This work was partly conducted 
at the Jet Propulsion Laboratory, California Institute of Technology, under contract with the National Aeronautics and 
Space Administration (NASA). This paper is partly based on observations made with the {\it XMM-Newton} Observatory, a European 
Space Agency (ESA) science mission with instruments and contributions funded by ESA member states and the USA. Some of the data 
presented herein were obtained at the W.M. Keck Observatory, which is operated as a scientific partnership among the California 
Institute of Technology, the University of California and NASA. The Observatory was made possible by the generous financial 
support of the W.M. Keck Foundation. We have made use of the NASA/IPAC Extragalactic Database (NED) which is operated 
by the Jet Propulsion Laboratory, California Institute of Technology, under contract with NASA. 


\vfill
\eject

\begin{deluxetable}{ccccc}
\tablecaption{3C 120 SED.}
\tablewidth{0pt}
\tablehead{
\colhead{Waveband} &
\colhead{$\log \nu $ (Hz)} & \colhead{$\log E$ (eV)} & \colhead{$\log \nu F_\nu$ (W m$^{-2}$)} 
& \colhead{$\log \nu F_{\nu,\mathrm{c}}$\tablenotemark{a} (W m$^{-2}$)}}
 
\startdata
37 GHz    & 10.57  & -3.82 & -15.12 & \nodata \\
250 GHz   & 11.40 & -2.99 & -13.99  & \nodata \\
V     & 14.74 &  0.35 & -13.54 & -13.15 \\
UVW1  & 15.01 &  0.63 & -13.18 & -12.44 \\
1 keV & 17.38 &  3.00 & -13.73 & -13.55 \\
2 keV & 17.68 &  3.30 & -13.63 & -13.60 \\
\enddata

\tablenotetext{a}{Flux values corrected for Galactic absorption.}
\end{deluxetable}

\begin{deluxetable}{ccccc}
\tablecaption{3C 120 optical emission lines.}
\tablewidth{0pt}
\tablehead{
\colhead{Line} & \colhead{$\lambda_\mathrm{p}$ (\AA ) } & \colhead{$v$ (km s$^{-1}$)} & \colhead{FWHM(km s$^{-1}$)} & 
                  \colhead{F\tablenotemark{a}($10^{-14}$ erg s$^{-1}$ cm$^{-2}$)}}
 
\startdata
H$_\epsilon$  & 3970.07 & \nodata        & \nodata        & 10 $\pm$ 2  \\
H$_\delta$(n)    & 4101.73 & \nodata     & \nodata        & 13 $\pm$ 1  \\
H$_\gamma$(n) & 4340.46 & \nodata        & \nodata        & 20 $\pm$ 2  \\
H$_\gamma$(b1)& 4340.46 & \nodata        & \nodata        & 30 $\pm$ 2  \\
He {\sc ii}   & 4686    &  210 $\pm$ 170 & 6300 $\pm$ 400 & 43 $\pm$ 2  \\
H$_\beta$(n)  & 4861.32 & -110 $\pm$ 20  & 1970 $\pm$ 30  & 50 $\pm$ 2  \\
H$_\beta$(b1) & 4861.32 & 1070 $\pm$ 90  & 5700 $\pm$ 100 & 59 $\pm$ 3  \\
H$_\beta$(b2) & 4861.32 & 5600 $\pm$ 300 &13,700 $\pm$ 600& 76 $\pm$ 3  \\
$[$O III$]$   & 4959    &   \nodata      & \nodata        & 17 $\pm$ 1  \\
$[$O III$]$   & 5007    & -283 $\pm$ 7   &  560 $\pm$ 20  & 61 $\pm$ 2  \\
$[$Fe {\sc vii}$]$& 5159& -210 $\pm$ 70  &  330           & 2.0$\pm$0.4 \\
He {\sc i}    & 5876    &  200 $\pm$ 60  & 2700 $\pm$ 200 & 16 $\pm$ 1  \\
$[$Fe {\sc vii}$]$ & 6087 & \nodata      & \nodata        & 0.8$\pm$0.3 \\
$[$O {\sc i}$]$ & 6364  & -280 $\pm$ 110 &  450 $\pm$ 170 & 0.9$\pm$0.3 \\
H$_\alpha$(n) & 6562.8  &   \nodata      & \nodata        & 243$\pm$ 5  \\
H$_\alpha$(b1)& 6562.8  &   \nodata      & \nodata        & 206$\pm$ 5  \\
H$_\alpha$(b2)& 6562.8  &   \nodata      & \nodata        &  36$\pm$ 3  \\
$[$S {\sc ii}$]$ & 6716 &    0           &  330 $\pm$ 40  & 3.0$\pm$0.3 \\
$[$S {\sc ii}$]$ & 6731 &    0           & \nodata        & 2.7$\pm$0.4 \\
He {\sc i}    & 7065    &   \nodata      & \nodata        & 9.6$\pm$0.8 \\
$[$Ar {\sc iii}$]$ & 7136 &-30 $\pm$ 90  &  830           & 2.4$\pm$0.4 \\
He {\sc i}   & 8447     &   \nodata      & \nodata        &11.5$\pm$0.7 \\

\enddata
\tablenotetext{a}{Flux values corrected for Galactic extinction.}
\end{deluxetable}

\vfill
\eject

\end{document}